**Ultrastrong Coupling of Band-Nested Excitons in Few-Layer Molybdenum Disulphide**

*Aaron H. Rose, Taylor J. Aubry, Hanyu Zhang, Jao van de Lagemaat\**

A. H. Rose, T. J. Aubry, H. Zhang, J. van de Lagemaat
National Renewable Energy Laboratory
Chemistry and Nanoscience Center
15013 Denver W Pkwy
Golden, CO 80401, USA
E-mail: Aaron.Rose@nrel.gov, Jao.vandeLagemaat@nrel.gov

H. Zhang (current)
First Solar
1035 Walsh Ave
Santa Clara, CA 95050, USA

The two-dimensional transition-metal dichalcogenides (2D TMDCs) are an intriguing platform for studying light–matter interactions because they combine the electronic properties of conventional semiconductors with the optical resonances found in organic systems. However, the coupling strengths demonstrated in strong exciton–polariton coupling remain much lower than those found in organic systems. In this paper, we take on a new approach by utilizing the large oscillator strength of the above-band gap C exciton in few-layer molybdenum disulphide (FL-$MoS_2$). We show a *k*-space Rabi splitting of 293 meV when coupling FL-$MoS_2$ C excitons to surface plasmon polaritons at room temperature. This value is 11% of the uncoupled exciton energy (2.67 eV or 464 nm), ~2× what is typically seen in the TMDCs, placing the system in the ultrastrong coupling regime. Our results take a step towards finally achieving the efficient quantum coherent processes of ultrastrong coupling in a CMOS-compatible system—the 2D TMDCs—in the visible spectrum.

1. Introduction

1.1. Strong coupling and the 2D TMDCs

Two-dimensional (2D) transition-metal dichalcogenides (TMDCs) have intriguing optical, electronic, and catalytic properties.[1–3] Their absorption is characterized by sharp excitonic resonances, making the TMDCs good candidates for strong exciton–polariton coupling.[4] In fact, one of the most exciting results in exciton–polariton systems—polariton condensation and lasing[5–7]—has recently been demonstrated in the TMDCs.[8–10] However, coupling strengths have remained much lower than what is achievable in organic systems, with typical Rabi splitting values of ≲ 150 meV or ≲ 6% of the exciton energy. In organic systems, Rabi splitting of >10%, which is in the ultrastrong coupling regime, is common.[11–18] In the ultrastrong coupling regime, strong coupling interactions are expected to be more efficient and emergent quantum coherent phenomena such as ground-state virtual photons and entangled pairs are predicted to exist even at room temperature or without resolvable Rabi splitting due to broad resonators.[19,20] As such, it is desirable to achieve this regime in a CMOS-compatible system like the TMDCs. In this paper, we exploit the large oscillator strength of the C exciton in few-layer molybdenum disulphide (FL-$MoS_2$) to achieve ultrastrong coupling at room temperature.

1.2. The C exciton of 2D TMDCs

In the 2D TMDCs, near the band edge, the two lowest energy exciton transitions are called the A and B excitons. The A and B exciton absorption features in FL-$MoS_2$ are labelled in **Figure 1a**. These excitons correspond to transitions between the conduction band minimum and the spin–orbit-split valence band maxima at the K point in **Figure 1b**. The next prominent absorption feature is the C exciton, which has significantly larger oscillator strength than the A or B, as seen by comparing the Lorentzian fits in Figure 1a.

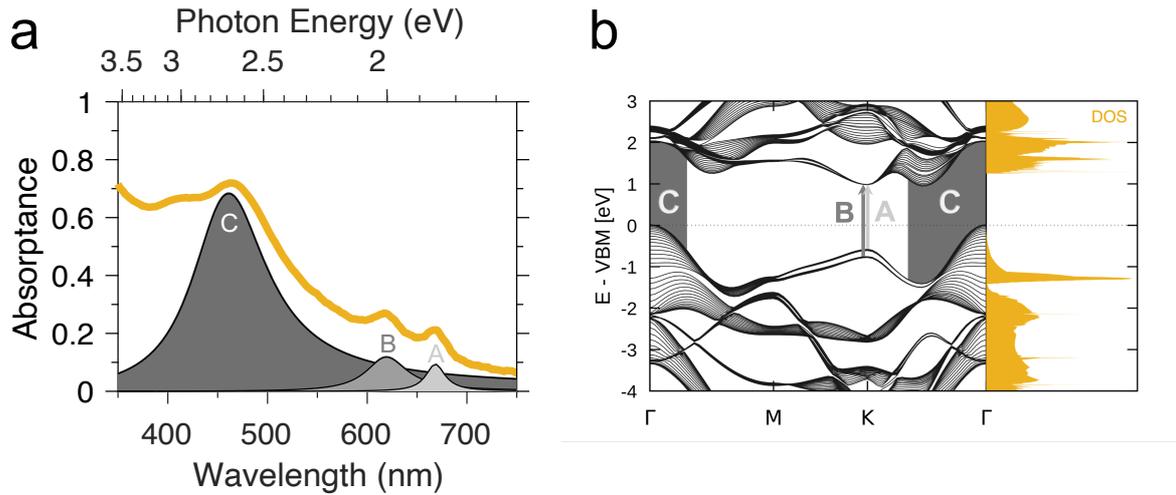

**Figure 1.** a) Absorptance, $A$, of FL-MoS$_2$ film on 40 nm Ag on glass. Reflectance, $R$, of film was taken and $A$ calculated as $A = 1 - R$, where transmission is assumed to be negligible. b) Band structure and density of states of FL-MoS$_2$ (17 layers), predicted from density functional theory. The arrows labeled A and B point out A and B exciton transitions while the shaded region labelled C shows the nested band region where C exciton transitions may be excited.

Interestingly, the C exciton does not arise from transitions between two parabolic bands, but rather between regions of parallel bands, referred to as nested bands, shown by the shaded region in Figure 1b.[21] The C exciton oscillator strength is so large because the region of $k$-space with parallel bands, and thus the density of states in this region, is so large.

An unusual consequence of the nested bands is that C exciton carriers are expected to spontaneously self-separate in momentum space[22–24] and exhibit slowed hot carrier cooling relative to the A and B excitons.[23] This has been seen by Wang et al. and ourselves in ultrafast spectroscopic studies.[23,25] Wang et al. coupled MoS$_2$ to graphene and calculated 80% charge extraction efficiency from the C exciton, showing the potential for harvesting hot carriers. We showed that the ultrastrong coupling analyzed here leads to increases in the lifetimes of the two slowest C exciton decay processes by factors of 1.5 and 5.8 and so may be beneficial to aiding hot carrier extraction from the C exciton in the TMDCs.

Another intriguing demonstration of C exciton physics in the TMDCs is the 400-fold enhancement of second harmonic generation (SHG) from the C exciton in monolayer WS$_2$ when weakly coupled to surface plasmons.[26] Strong coupling has also been used to enhance SHG in organics and TMDCs while also modifying the emission spectra via Rabi splitting.[27,28] Thus, strong coupling may offer a way to further tune and enhance the SHG, as well as other non-linear properties, at the C exciton in the TMDCs.

2. Analysis in $k$-space

2.1. Experimental and simulated exciton–plasmon dispersion

We use the Kretschmann-Raether technique shown in **Figure 2a** to measure the p-polarized angle-resolved reflectance spectrum, $R(\theta)$, and calculate the experimental dispersion of **Figure 2b**. Absorptance, $A(\theta)$, is calculated as $A(\theta) = 1 - R(\theta)$, where transmission is negligible as we are under conditions of total internal reflection. The 40 nm Ag supports propagating surface plasmons that are excited through the backside of the prism (see Supporting Information (SI) Figure S2f for an example of the plasmon dispersion in the absence of strong coupling). We transfer FL-MoS$_2$ grown by chemical vapor deposition onto the Ag to complete the samples. Finally, we convert the angle-space $A(\theta)$ dispersion into the more physically relevant wavevector-space $A(k_{//})$ by $k_{//} = (2\pi/\lambda)n \sin \theta_i$, where $n$ is the wavelength-dependent lossless refractive index of the prism and $\theta_i$ is the angle of incidence within the glass prism.

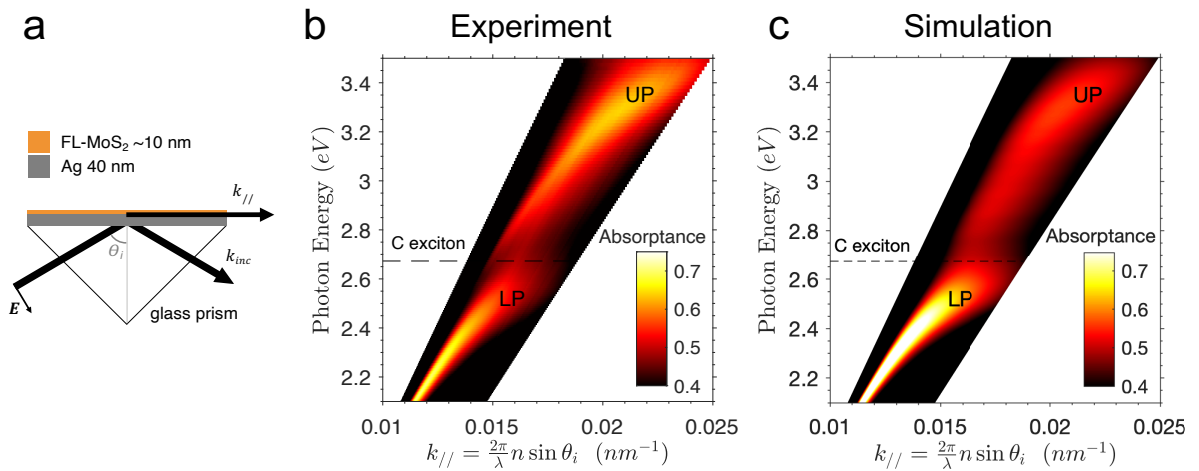

**Figure 2.** a) Kretschmann-Raether method of measuring angle-resolved dispersion. P-polarized light is incidence on the Ag–MoS$_2$ film through the bottom of the prism. The surface plasmon polariton propagates in the direction of $k_{//}$. b) Experimental dispersion. c) Transfer matrix model of the experimental dispersion. In (b) and (c), the upper polariton (UP) and lower polariton (LP) branches are labeled while the dashed line indicates the position of the uncoupled C exciton energy.

Compared to our earlier work showing strong coupling at the A and B excitons in thicker MoS$_2$,[29] the features near the A and B excitons show only slight perturbations to the plasmon dispersion, indicating minimal plasmon coupling to these excitons (see SI Figure S1 for dispersion showing the A and B excitons near 1.85 and 2.0 eV). However, at the C exciton position of 2.67 eV, a large Rabi splitting, or avoided crossing, is observed, indicating strong coupling. There is minimal overlap of two coupled modes labeled upper polariton(UP) and lower polariton(LP) due to the large splitting. We characterize the magnitude of the splitting in the next section.

In order to derive inputs for the semi-classical coupled harmonic oscillator model used in the next section we model the observations using a transfer matrix model. The transfer matrix simulation results are shown in **Figure 2c**. The simulated system reproduces the experimental data well and thus can be used to design future studies, e.g. photoelectrochemistry studies in liquid electrolyte, which we know shifts the plasmon and requires judicious choice of optical glass for the prism. The model was built in MATLAB[30] as NBK7 substrate / 1 nm Ti / 43 nm Ag / 0.27 nm Ag$_2$S / 8 nm MoS$_2$. The substrate refractive index and thicknesses of Ag and MoS$_2$ were fit to reproduce the data. The NBK7 refractive index fit matches specifications of Schott within ~2 %. The fit value for the MoS$_2$ thickness of 8 nm is near the value measured by ellipsometry of 10 nm. The refractive index for MoS$_2$ was measured from the top of the prism by ellipsometry and is shown in Figure S2a. Refractive indices for Ag were modeled from ellipsometry data on witness samples and are in-line with those of Palik[31] and Johnson and Christy[32] and the fit thickness was within 2 nm of that measured with ellipsometry. Refractive indices for Ti and Ag$_2$S were taken from refs [33] and [34] respectively. The thickness of Ti was set to be 1 nm while the thickness of Ag$_2$S was derived from ellipsometry.

2.2. Coupled oscillator model

The transfer matrix simulations serve to verify that our observations are to be expected based on applying Maxwell's equations to the system at hand. However, they do not quantify the strength of the coupling or the composition of the new hybrid modes. To gain this insight, we solve the problem using a semiclassical coupled oscillator model. The Hamiltonian of the system can be written as **Equation 1**:

$$\begin{bmatrix} E_{SPP}(k_{//}) & \frac{\Omega_A}{2} & \frac{\Omega_B}{2} & \frac{\Omega_C}{2} \\ \frac{\Omega_A}{2} & E_A & 0 & 0 \\ \frac{\Omega_B}{2} & 0 & E_B & 0 \\ \frac{\Omega_C}{2} & 0 & 0 & E_C \end{bmatrix} \quad (1)$$

In this model, the problem is treated as a four-level quantum system coupled to classical electromagnetic fields. $E_{SPP}(k_{//})$ is the uncoupled or "bare" plasmon energy, which is a function of in-plane wave vector as given by its dispersion (discussed below). $E_A$, $E_B$, and $E_C$ are the bare exciton energies. $\Omega_A$, $\Omega_B$, and $\Omega_C$ are the Rabi splitting values. The off-diagonal zeros represent that there is no coupling between excitons, which is expected to be the case. The Hamiltonian is solved and the eigenvalues give the four coupled modes, shown as the yellow lines in **Figure 3**. The model is fit by the least squares routine at each $k$-point to the experimental data. The experimental data, plotted in black in Figure 3, is represented by the peak energies found from fitting each branch of the dispersion to Lorentzians at each $k$-point.

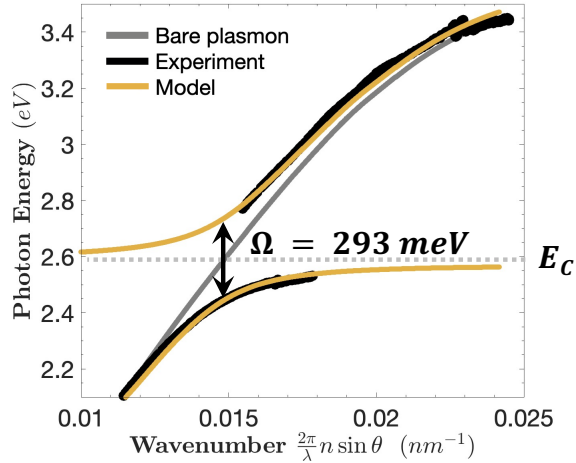

**Figure 3.** Coupled oscillator model fit to data. Black data points are from peaks in experimental UP and LP data of Figure 2b. The gray line is the simulated bare plasmon, $E_{SPP}(k_{//})$. Yellow lines are coupled oscillator model fits to the data. The gray dotted line is a fit to the C exciton energy position. The best fit value of the Rabi splitting is $\Omega$ = 293 meV or 11% of splitting energy, indicating that we are in the strong coupling regime.

To do the fitting, $E_A$, $E_B$, and $E_C$ are set as fit parameters, with initial values taken from the positions of the absorptance peaks in Figure 1a. $\Omega_A$, $\Omega_B$, and $\Omega_C$ are also set as fit parameters. Initial values of $\Omega$ are found by subtracting the energy of the branches on either side of each exciton and taking the minimum difference. $E_{SPP}(k_{//})$ is not set as a fitting parameter, but is fed into the model.

Determining the bare plasmon energy, $E_{SPP}(k_{//})$, shown by the gray line in Figure 3, is more complicated than simply measuring the dispersion of a control Ag sample with no $MoS_2$. Plasmons are sensitive to their local environment, defined by the near-field of the electromagnetic plasmon wave, ~10–100 nm. Therefore, the presence of several nanometers of any material, lossy or not, will modify the plasmon dispersion. In this case with $MoS_2$ on Ag, the plasmon asymptotically approaches ~3.5 eV, which is red-shifted from the bare Ag case. In solving the similar problem in our previous work,[29] the bare plasmon was approximated by solving the transfer matrix model of the system with lossless $MoS_2$. The lossless $MoS_2$ was modeled by setting the imaginary part of the complex $MoS_2$ refractive index to zero. In the case here, however, that approach had to be modified. Due to the Kramers-Kronig relationship, the real and imaginary parts of the refractive index are coupled such that a peak or valley in one corresponds to an inflection point in the other. Therefore, setting the imaginary part of the refractive index to zero to model a lossless version of that material is not physically accurate, as features are maintained in the real part that should disappear. We show the refractive index and modeled plasmon using this method in Figure S2c,d. In the previous work, the perturbations at the A and B excitons obtained using this method were negligible. Here, due to the large oscillator strength of the C exciton, the perturbations are too large to be ignored. Instead, the fictitious lossless $MoS_2$ must be modeled with a smoothly-varying real part of the refractive index. We fit the refractive index using a Cauchy+Urbach absorption model with J.A. Woollam WVASE software, fitting only the plasmonic extremes of the experimental dispersion data. The resulting refractive index and modeled lossless plasmon are show in Figure S2e,f. This dispersion is then fit to a Lorentzian at every $k$-point and the peak positions comprise $E_{SPP}(k_{//})$.

2.3. Coupled oscillator model Hopfield coefficients

The new hybrid modes, $|\phi\rangle_k$, may be written in the basis of the uncoupled modes according to **Equation 2**.

$$|\phi\rangle_k = c_1|SPP\rangle + c_2|A\rangle + c_3|B\rangle + c_4|C\rangle \qquad (2)$$

The Hopfield coefficients[35] $|c_1|^2$ represent the plasmonic and excitonic (A, B, or C) weighting of the hybrid modes, as a function of wavevector. **Figure 4** plots these coefficients for each mode.

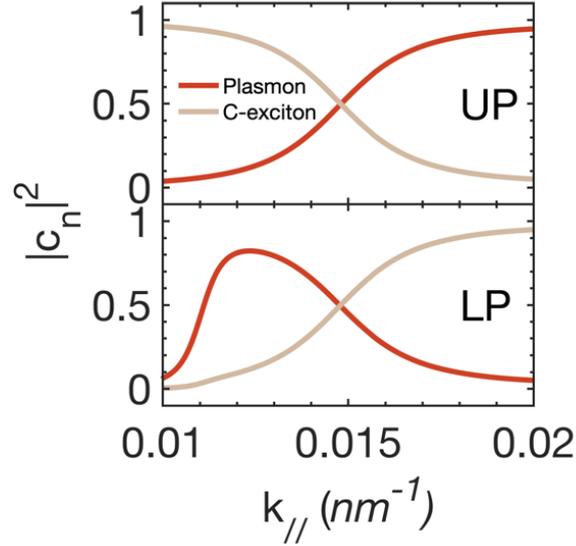

**Figure 4.** Hopfield/weighting coefficients of coupled modes in terms of uncoupled modes, taken from semiclassical oscillator model fit to experimental data. Here we just focus on the two branches, UP and LP, on either side of the C exciton. See Figure S4 for the coefficients of all four branches.

The analysis shows the expected result that at the crossing point of the uncoupled modes of Figure 3, the weighting is a 50-50 exciton–plasmon hybrid. In Figure S4 we see again that there is some hybridization between the A and B excitons, as we saw in our earlier work focusing on strong coupling to them. Here though, there is no evidence that the C exciton is hybridized w/ the A and B excitons, as the energy difference between them is too large.

2.4. Discussion

The resulting best fit values for the bare exciton energies and Rabi splitting values are summed up in **Table 1**. The experimental bare exciton energies and Rabi splitting values are also shown. The difference between the experimental and fit values of the C exciton energy and Rabi splitting are ≤ 1%, indicating a good fit to the model in this region. The most important finding is that the Rabi splitting at the C exciton is 293 meV, which is $\eta = \frac{\Omega_C}{E_C} = 11\%$ of the uncoupled exciton energy. Compared to ~70 meV splittings at the A and B excitons in similar systems, this represents a factor of 2.4 increase in $\eta$, and enters the ultrastrong coupling regime ($\eta > 10\%$). We further note that the same analysis performed in $\theta$-space shows Rabi splitting of 532 meV which gives $\eta = 21\%$ (see Figures S5-8 and Table S1).

**Table 1.** Best fit parameters from coupled oscillator model for strong coupling to the C exciton in MoS$_2$. Solved in $k$-space.

|  | $E_A$ (eV) | $E_B$ (eV) | $E_C$ (eV) | $\Omega_A$ (meV) | $\Omega_B$ (meV) | $\Omega_C$ (meV) | $\eta = \dfrac{\Omega_C}{E_C}$ |
|---|---|---|---|---|---|---|---|
| Experiment | 1.86 | 2.01 | 2.67 | 58 | 77 | 296 | 11% |
| Model | 1.86 | 2.01 | 2.59 | 67 | 95 | 293 | 11% |

One of the first papers demonstrating ultrastrong coupling used an organic J-aggregate dye molecule coupled to a plasmonic nanohole array to achieve 250 meV Rabi splitting at 1.8 eV transition energy, a ratio of $\eta = 14\%$.[11] In 2009, $\eta = 20\%$ was reached with J-aggregates coupled to a plasmonic nanodisk array.[12] This was pushed to $\eta = 32\%$ in another organic molecule, spiropyran, coupled to a dielectric cavity mode.[13] In 2014, the organic semiconductor squaraine was coupled again to a microcavity to achieve $\eta = 60\%$.[14] In inorganic semiconductors, the ultrastrong coupling regime was first seen in 2009 by coupling microcavity modes to intersubband transitions (transition energy of ~100 meV) in GaAs to achieve $\eta = 44\%$.[36] The only demonstration of ultrastrong coupling at optical frequencies in the TMDCs was in many-layer WS$_2$ at cryogenic temperatures. The Fabry-Perot resonance of a 32 nm flake was coupled to the A exciton to achieve 270 meV, or $\eta = 13\%$.[37] There are many other reports of $\eta = 4 - 9\%$ in the TMDCs.[4,38–42] All of these reports couple to the lowest energy A or B excitons. By targeting the C exciton, we are able to achieve ultrastrong coupling at room temperature in the TMDCs.

3. Conclusions

In summary, we find ultrastrong exciton–polariton coupling at the C exciton in FL-MoS$_2$, at room temperature. The observed Rabi splitting of 293 meV ($\eta = 11$ %) is significantly larger than that found in previous studies using TMDCs, due to the large C exciton oscillator strength.

In the ultrastrong coupling regime, quantum coherent phenomena such as ground state virtual photons becomes efficient.[19,43] Another application of strong coupling is to modify the chemical behavior of materials. Strong coupling results in shifted band positions or energy levels, which can lead to fundamentally new electronic and catalytic properties. Thus, there has been some work showing that chemical photoreaction pathways and rates can be manipulated by strongly coupling the reactants or catalysts with polariton modes.[44–48] Thus, ultrastrong coupling to the C exciton in TMDCs may find use in charge- or energy-transfer mediated photocatalytic chemical reactions, where the C exciton may have favorable energy and band-alignment for coupling to molecular orbitals of reactive species, intermediates, or products. Further applications specific to the C exciton in the TMDCs include slowed hot carrier cooling for enhanced conversion of solar energy into electricity or chemicals and enhanced SHG for coherent light sources in the violet/blue.

Our findings, therefore, suggest that the C exciton of the TMDCs is a suitable platform for studying ultrastrong coupling physics in a CMOS compatible inorganic semiconductor at room temperature.

4. Experimental Section/Methods

*Theoretical Calculations*
The band structure of Figure 1b was calculated using ab initio density functional theory (DFT) methods to verify the band-nested structure in our 10 nm FL-$MoS_2$ samples. Although standard correlation functionals such as GGA–PBE used here tend to underestimate the band gap, we expect the band profile to remain accurate. In comparing to previous G0W0 calculations on mono-to-few layer (1–6) $MoS_2$,[49] we find that the band profile remains similar. We calculated 16- and 17-layer structures to both span the 10 nm thickness of interest (calculated thicknesses of 9.7–10.3) and account for any even and odd layer dependence in the band structure.[50]

All calculations were performed using DFT with the JDFTx24 software implementation. The Perdew–Burke–Ernzerhof (PBE) form of the generalized gradient approximation was used to describe the exchange−correlation interaction. The valence electron–nuclear interactions were described by optimized fully-relativistic norm-conserving Vanderbilt pseudopotentials (ONCVPSP)[51] from the PseudoDojo Project[52] (stringent accuracy). A 45 Hartree cut-off was used for the plane-wave basis set. Calculations were performed with fully-relativistic spin to capture the effects of spin–orbit coupling on the MoS$_2$ band-structure. To capture the van der Waals interlayer interactions, the Grimme DFT + D2 scheme was used.[53]

The 16- and 17-layer MoS$_2$ slabs were constructed from lattice optimized bulk MoS$_2$. To prevent interaction between images, we employed a vacuum layer of at least 15 Å and truncated coulomb potentials[54] in the out-of-plane direction. The 2D-multilayers were also allowed to fully relax to a final lattice parameter of 3.192 Å for both 16- and 17-layer structures and a final thickness of 9.7 nm and 10.3 nm, respectively. To sample the Brillouin zone, a Γ-centered k-point sampling of 12 × 12 × 3 was used for the bulk and 12 x 12 x 1 for the 2D-multilayers. The convergence criteria were set to 1x10−6 Hartree for both structure and energy optimizations and converged via a variational minimization algorithm.[55] Finally, the density-of-states was obtained using the tetrahedron method for Brillouin zone integration.[56]

*Prism preparation*

The right-angle prism used as a substrate was purchased from Edmund Optics (stock number 32-334), made from uncoated N-BK7 glass, with 20 mm leg length. It was cleaned by sonicating in acetone then isopropyl alcohol for several minutes, soaking in an aqua regia bath (3:1 hydrochloric acid:nitric acid) for an hour, rinsing twice with Milli-Q water, drying with a nitrogen spray gun, and further desiccating on a hot plate at 150 °C for several minutes. The cleaned prism was stored in a nitrogen dry box between cleaning and Ag deposition. Ag deposition was performed by electron beam deposition of ~1 nm Ti followed by 40 nm of Ag, without breaking vacuum. The Ti serves as an adhesion layer but also damps the surface plasmon, so a minimal thickness was desired. The thickness of Ti/Ag was verified by profilometry and ellipsometry.

*Few-layer MoS$_2$ CVD growth*

The FL-MoS$_2$ growth procedures were adopted and modified from previous methods developed by Yu et al.[57] The chemical vapor deposition (CVD) growth was carried out by a three-temperature-zone furnace. 500 mg of sulfur pellets (Sigma Aldrich) were placed at Zone 1, while the sapphire wafer (University Wafer) was located at Zone 3. An insert tube in Zone 2 was used to create an isolated local environment. 2 mg of MoO$_3$ powder (Sigma Aldrich) was loaded into the insert tube. 25 sccm Ar/O$_2$ (4 vol. % of O$_2$) premix gas was flowed over the MoO$_3$ powder into Zone 2. The O$_2$ allows the MoO$_3$ precursor to remain in oxidized form as it is carried through the insert. 125 sccm of Ar was supplied at Zone 1 to carry the sulfur and balance the growth pressure at 1 Torr. During the growth, the temperatures in Zones 1, 2, and 3 were ramped at the rate of 35, 35, and 70 °C/min and then held for 30 min at 180, 530, and 930 °C, respectively.

*MoS$_2$ transfer*

To transfer the CVD MoS$_2$ film onto the prism, we followed the procedures described by Xu et al.[58] The CVD-grown MoS$_2$ was coated by polystyrene (PS, average molecular weight ~192,000, Sigma-Aldrich) dissolved in toluene (50 mg/ml) by spin-coating at 3,000 rpm for 60 seconds. The resulting PS/MoS$_2$/sapphire film was dried in a 150°C oven for 5 min. One edge of the film was scribed by a utility knife to create an opening for etchant to access the MoS$_2$/sapphire interface. The substrate was soaked in 80°C 2 M NaOH solution until the NaOH etched the interfacial sapphire across the wafer and the PS/MoS$_2$ film detached from the sapphire wafer. The PS/MoS$_2$ film was carefully transferred to an ultra-pure water (18.2 MΩ) bath to minimize NaOH residue. Then, the PS/MoS$_2$ film, which naturally floats on the water surface, was transferred to the prism by lifting the prism up from underneath the film. The PS coating was removed by soaking the prism in toluene. Finally, the prism was baked in N$_2$ at 300 °C for 30 min.

**Supporting Information**

Supporting Information is available from the Wiley Online Library or from the author.

Acknowledgements

This work was authored by the National Renewable Energy Laboratory, operated by Alliance for Sustainable Energy, LLC, for the U.S. Department of Energy (DOE) under Contract No. DE-AC36-08GO28308. Funding provided by U.S. Department of Energy Office of Science, Office of Basic Energy Sciences, Division of Chemical Sciences, Geosciences, and Biosciences, Solar Photochemistry Program. The views expressed in the article do not necessarily represent the views of the DOE or the U.S. Government.

Received: ((will be filled in by the editorial staff))
Revised: ((will be filled in by the editorial staff))
Published online: ((will be filled in by the editorial staff))

Table of Contents Entry

While strong coupling is common in inorganic systems, the ultrastrong regime, where quantum coherent phenomena become efficient, has been the domain of organic systems. Here, we exploit the large oscillator strength of the C exciton in $MoS_2$ to demonstrate ultrastrong coupling in a CMOS-compatible inorganic semiconductor in the visible spectrum at room temperature.

Aaron H. Rose, Taylor J. Aubry, Hanyu Zhang, Jao van de Lagemaat*

**Ultrastrong Coupling of Band-Nested Excitons in Few-Layer Molybdenum Disulphide**

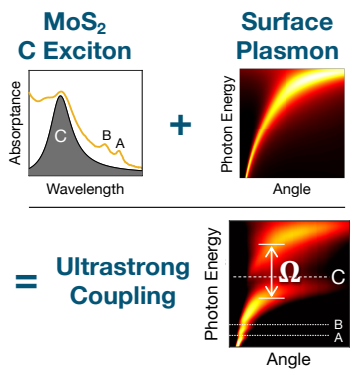

Supporting Information

**Ultrastrong Coupling of Band-Nested Excitons in Few-Layer Molybdenum Disulphide**

*Aaron H. Rose, Taylor J. Aubry, Hanyu Zhang, Jao van de Lagemaat\**

1. Additional data in $k$-space

**Figure S1** shows the experimental dispersion of Figure 2b for an extended range where the A and B exciton positions can be seen at 1.86 and 2.01 eV, respectively.

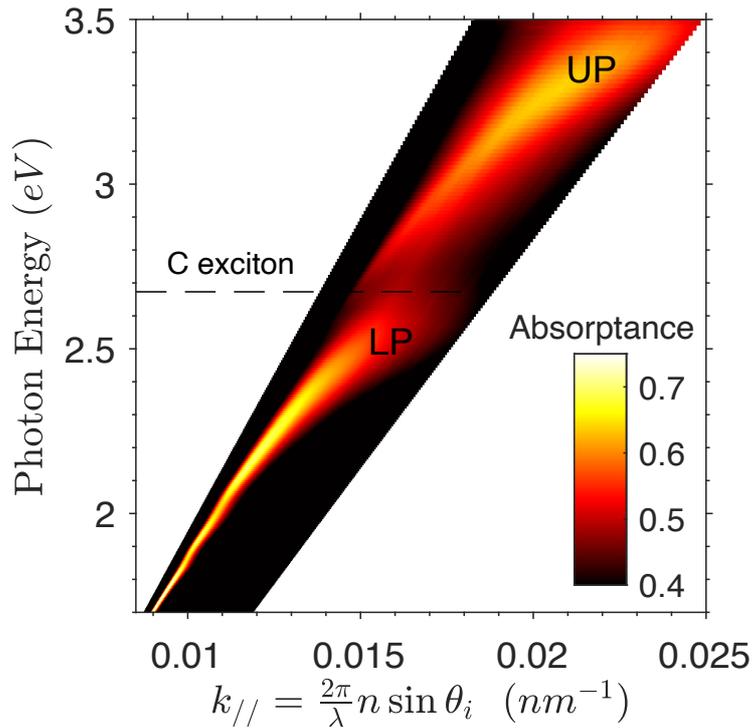

**Figure S1.** Experimental dispersion showing A and B exciton region. The upper polariton (UP) and lower polariton (LP) branches are labeled while the dashed line indicates the position of the uncoupled C exciton energy.

**Figure S2** shows the transfer matrix method (TMM) simulations of the experimental data (Figure S2b), poor approximation of the bare plasmon (Figure S2d), and good approximation of bare plasmon (Figure S2f) with their respective MoS$_2$ refractive indices (Figure S2a,c,e). Figures S2c,e show two methods of approximating the fictitious lossless MoS$_2$ needed to model the bare plasmon for the coupled harmonic oscillator model.

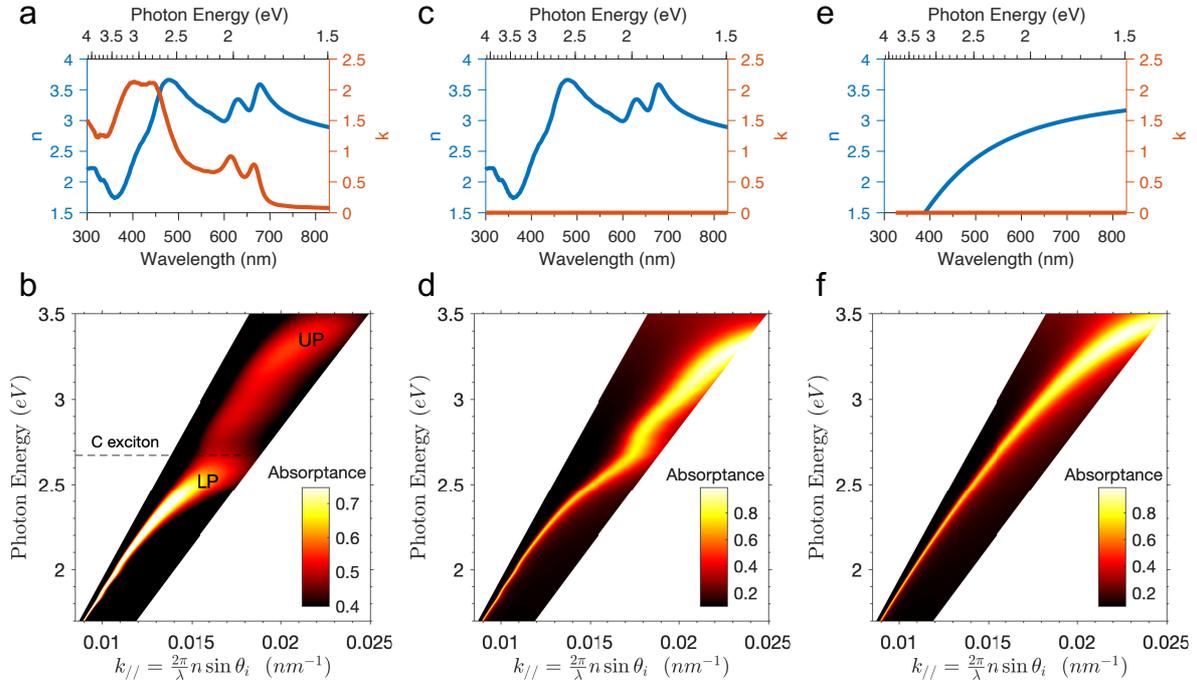

**Figure S2.** a) Real (n) and imaginary (k) parts of the FL-MoS$_2$ refractive index, measured by ellipsometry. b) TMM dispersion of experimental system, showing A and B exciton region. c) Refractive index of (a) with k set to zero. d) TMM with k = 0 refractive index, showing poor approximation of bare plasmon. e) Smoothly-varying refractive index model of lossless MoS$_2$. f) TMM with lossless MoS$_2$ from (e), showing typical Ag plasmon behavior, but redshifted to account for the thin film of MoS2 on top of the Ag. This plasmon dispersion fits the experimental data well at lower and higher energies, far away from the C exciton, where the coupled dispersion is nearly completely plasmonic.

**Figure S3** shows the coupled harmonic oscillator fit to the data of Figure 3 for a larger range showing the A and B exciton positions, in addition to the C exciton. While the fit to the dispersion data near the A and B excitons shows Rabi splitting, a look at Figure S1 shows that the splitting of the plasmon is not resolvable by eye, evidently showing that strong coupling isn't truly achieved in this region. However, in $\theta$-space shown below, we can see the region more closely and come to a different conclusion.

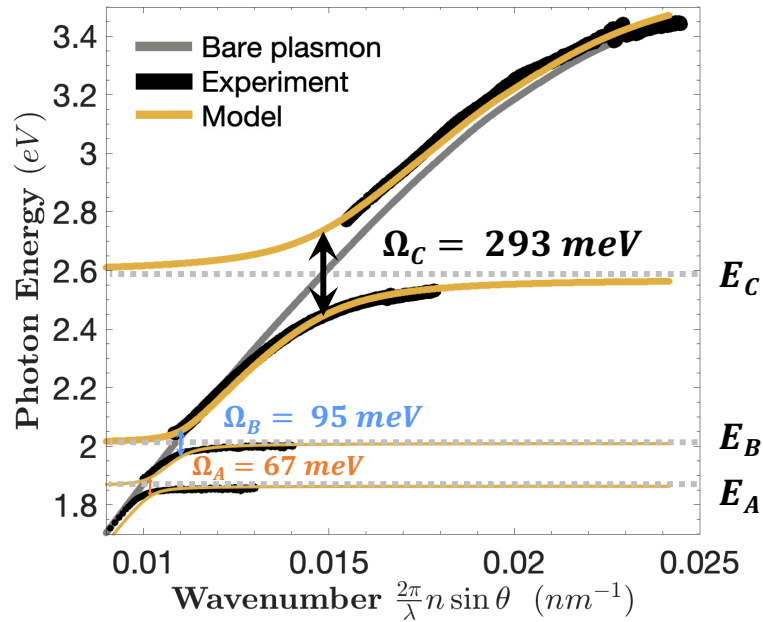

**Figure S3.** Coupled oscillator model fit to data. Black data points are from experiment. Gray line is simulated bare plasmon used in the model. Yellow lines are coupled oscillator model fit to the data. Gray dotted lines are model fit of A, B, and C exciton energy positions.

**Figure S4** shows the Hopfield coefficients for all four branches in the coupled oscillator model.

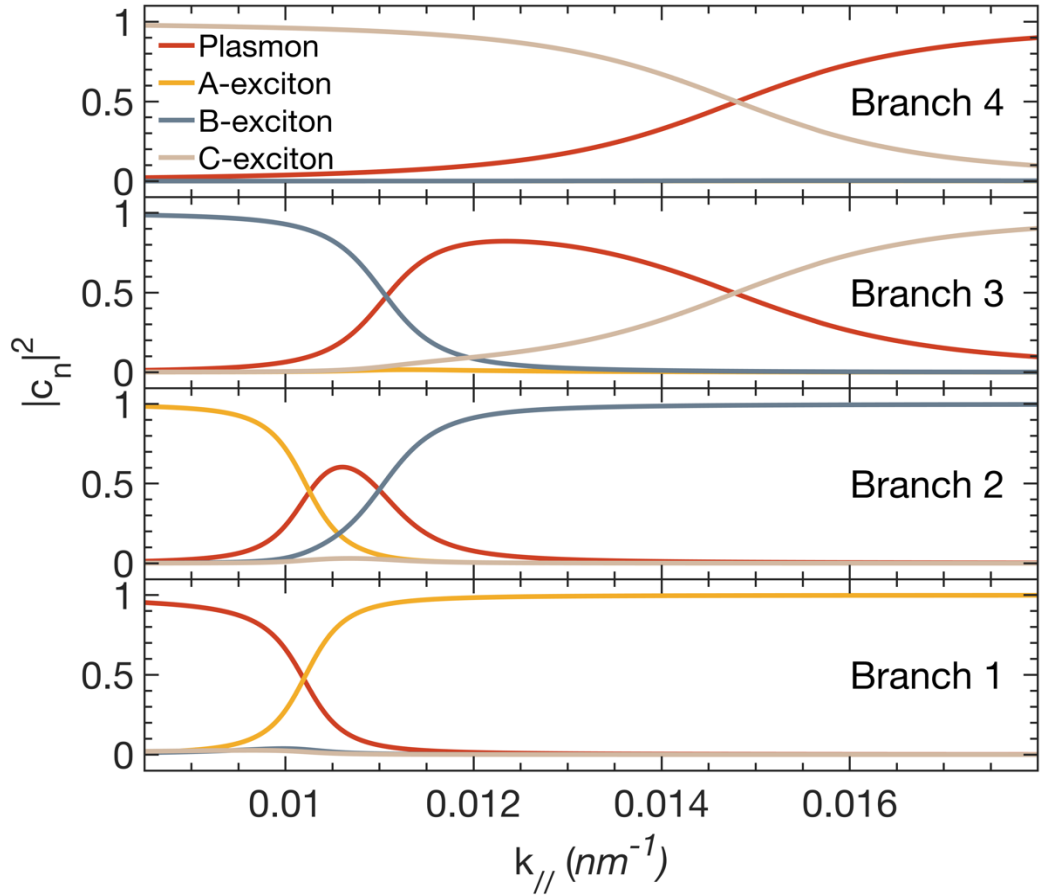

**Figure S4.** Hopfield/weighting coefficients of coupled modes in terms of uncoupled modes, taken from semiclassical oscillator model fit to experimental data. Branch 1 is the lowest energy branch seen in the dispersion in Figure S3, while Branch 4 is the highest branch. Coupling between excitons is marginal.

2. Analysis in $\theta$-space

The following figures show how the analysis unfolds if the reflectance/absorptivity vs. angle data is used without transformation into wavevector space. This analysis results in impressive Rabi splitting, but is not the appropriate basis to work in.[1] Still, it does show that at the point of minimum separation between coupled modes near the C exciton, the Rabi splitting seen experimentally is 561 meV, which corresponds to $\eta(\theta) = 21\%$ for C exciton transition energy of 2.67 eV.

**Figure S5** shows the experimentally simulated dispersion, showing the Rabi splitting very clearly.

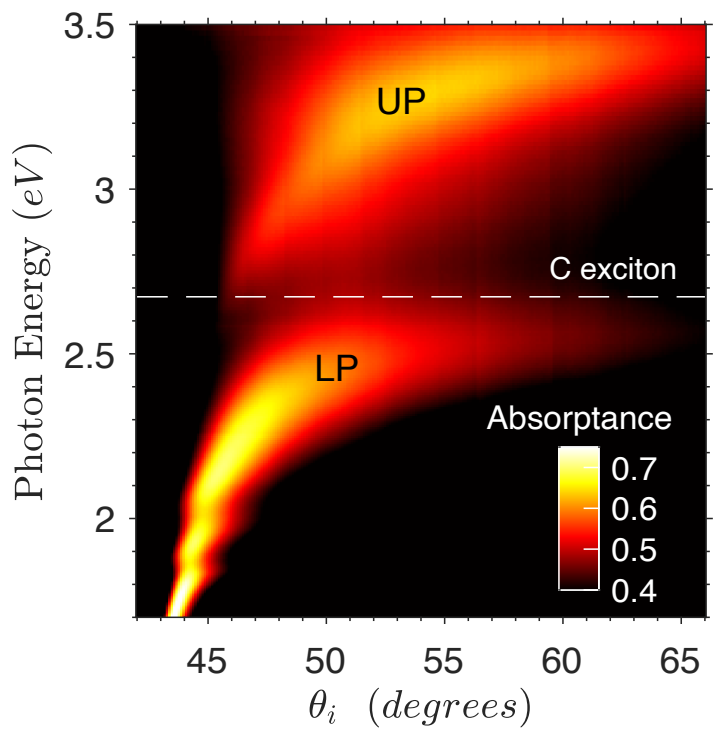

**Figure S5.** Experimental dispersion showing A and B exciton region. The upper polariton (UP) and lower polariton (LP) branches are labeled while the dashed line indicates the position of the uncoupled C exciton energy.

**Figure S6** reproduces Figure 2b,d,f in $\theta$-space.

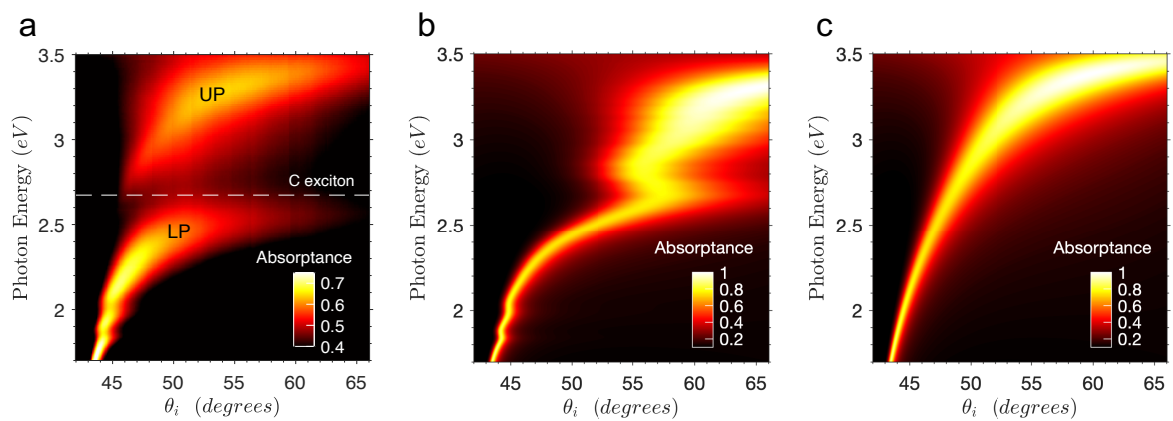

**Figure S6.** TMM simulation of the a) experimental dispersion, b) k = 0 bare plasmon, and c) smooth n bare plasmon, in $\theta$-space.

Applying the coupled oscillator model in $\theta$-space, we see from **Figure S7** that the Rabi splitting values are much larger than in $k$-space.

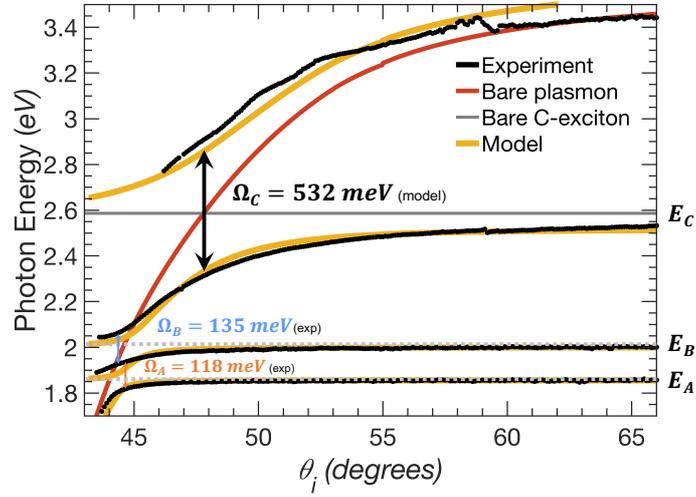

**Figure S7.** Coupled oscillator model fit to data in angle-space. Black data points are from experiment. Red line is simulated uncoupled plasmon used in the model. Yellow lines are coupled oscillator model fit to the data. Gray lines/dotted lines are model fit of A, B, and C exciton energy positions.

The best fit parameters from the coupled oscillator model in $\theta$-space are presented in **Table S2**. Again, the model fits Rabi splitting values to the A and B excitons. However, unlike in $k$-space, here the splitting at these excitons is resolvable by eye when looking at the experimental dispersion of Figure S5. Thus, there is in fact conventional strong coupling at the A and B excitons.

**Table S2.** Best fit parameters from coupled oscillator model for strong coupling to the C exciton in MoS$_2$ in $\theta$-space.

|            | $E_A$ (eV) | $E_B$ (eV) | $E_C$ (eV) | $\Omega_A$ (meV) | $\Omega_B$ (meV) | $\Omega_C$ (meV) | $\eta = \dfrac{\Omega_C}{E_C}$ |
|---|---|---|---|---|---|---|---|
| Experiment | 1.86 | 2.01 | 2.67 | 118 | 135 | 561 | 21% |
| Model      | 1.86 | 2.00 | 2.59 | 101 | 151 | 532 | 21% |

The Hopfield coefficients, shown in **Figure S8** show similar behavior as in *k*-space, with an asymmetry due to the $\theta$-space presentation.

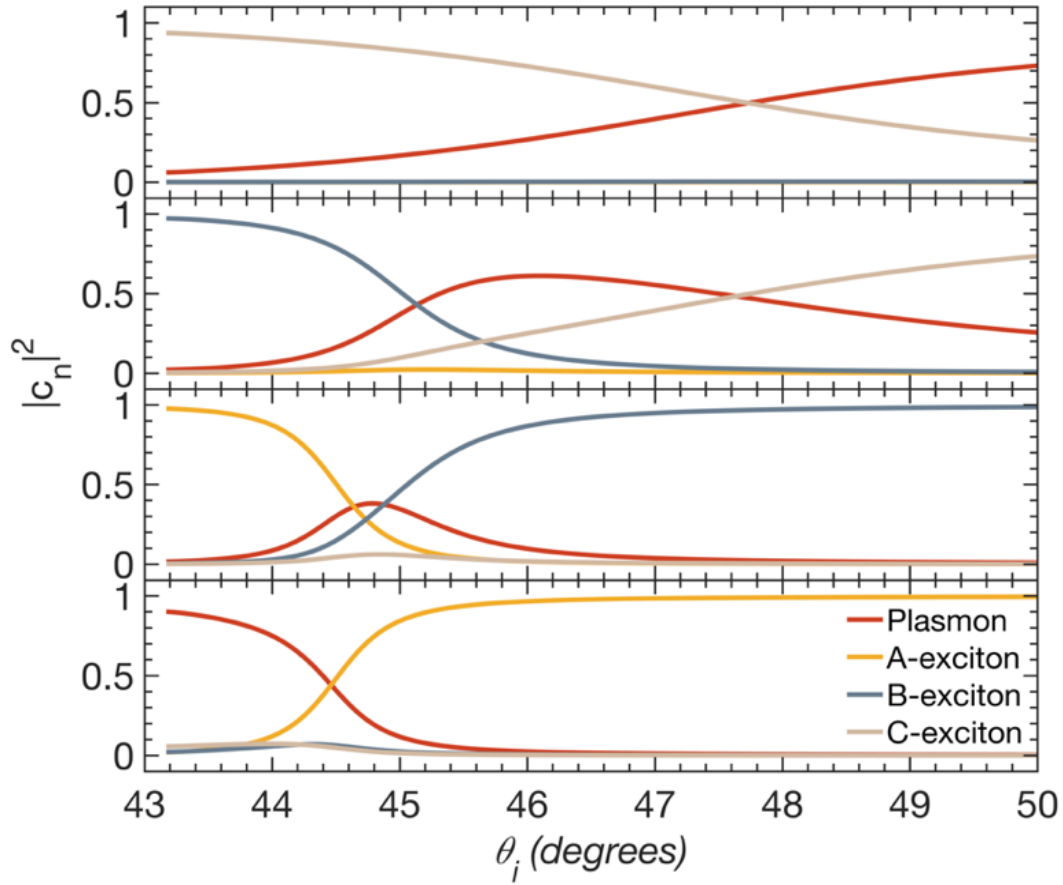

**Figure S8.** Hopfield/weighting coefficients of coupled modes in terms of uncoupled modes, taken from semiclassical oscillator model fit to experimental data, in $\theta$-space. Branch 1 is the lowest energy branch seen in the dispersion in Figure S7 while Branch 4 is the highest branch. Coupling between excitons is marginal. Asymmetry of curves is due to angle-space representation.

Supplementary Reference